\documentclass[twocolumn,aps,pre,superscriptaddress,showpacs,floatfix]{revtex4-1}
\usepackage[pdfborder=0 0 0]{hyperref}
\usepackage[dvips]{graphicx}
\usepackage{amsmath}
\usepackage{color}
\usepackage{amsfonts}
\usepackage{amssymb}
\usepackage{epstopdf}
\usepackage{colortbl}
\DeclareGraphicsExtensions{.eps,.png,.pdf}
\begin{document}
\markboth{Sethia}{Sethia}
%\title{\textit{Chimera} states in a globally coupled oscillator system}
\title{\textit{Chimera} states: the existence criteria revisited}
\author{Gautam C Sethia}
\email[e-mail: ]{gautam.sethia@gmail.com}
\affiliation{Institute for Plasma Research, Bhat, Gandhinagar 382 428, India}
\affiliation{Max-Planck-Institute for Physics of Complex Systems, 01187 Dresden, Germany}
\author{Abhijit Sen}
\affiliation{Institute for Plasma Research, Bhat, Gandhinagar 382 428, India}
\newcommand{\beq}{\begin{equation}}
\newcommand{\eeq}{\end{equation}}
\newcommand{\beqstar}{\[}
\newcommand{\eeqstar}{\]}
\newcommand{\bea}{\begin{eqnarray}}
\newcommand{\eea}{\end{eqnarray}}
\newcommand{\beastar}{\begin{eqnarray*}}
\newcommand{\eeastar}{\end{eqnarray*}}
\begin{abstract} 
{\it Chimera} states, representing a spontaneous break-up of a population of identical oscillators that are identically coupled, into sub-populations displaying
synchronized and desynchronized behaviour, have traditionally been found to exist in weakly coupled systems and with some form of nonlocal coupling between
the oscillators. Here we show that neither the weak-coupling approximation nor nonlocal coupling are essential conditions for their existence. We obtain for the first time
{\it amplitude-mediated chimera} states in a system of globally coupled complex Ginzburg-Landau oscillators. We delineate the dynamical origins for the formation of such states
from a bifurcation analysis of a reduced model equation and also discuss the practical implications of our discovery of this broader class of {\it chimera} states.
\end{abstract}

\maketitle
The spontaneous break-up of a system of identical oscillators, that are identically coupled, into sub-groups of oscillators with different synchronous properties is a fascinating collective phenomenon that was first reported by Kuramoto and Battogtokh\cite{kuramoto02} and has since been the subject of many investigations \cite{shima04,abrams04,kuramoto06,kawamura07,sethia08,omelchenko08,abrams08,sheeba09,laing09a,martens10,martens10a,ma10,bordyugov10,olmi10,lee11,wolfrum11a,wildie12,
laing12,gu13,panaggio13,zhu13b,lin13,omelchenko13a,omelchenko13b,omelchenko13c,panaggio13,singh13,ujjwal13,pazo14,zakharova14}. This spatio-temporal pattern of co-existing synchronous and de-synchronous oscillations, named as a {\it chimera} state by Abrams and Strogatz \cite{abrams04},  has also been experimentally demonstrated in a number of laboratory systems \cite{tinsley12,hagerstrom12,nkomo13,wickramasinghe13,martens13,larger13}. The natural manifestation of this state can be seen in such phenomena as unihemispherical sleep in many animals \cite{rattenborg00,mathews06} where the awake side of the brain shows desynchronized electrical activity, whereas the sleeping side is highly synchronized \cite{abrams08} or in the human brain when in certain regions the neuronal activity gets highly synchronized during epileptic seizures \cite{ayala73} or damage due to Parkinson's disease \cite{levy00}.  In model studies mentioned above {\it chimera} states have been found in phase only oscillator systems and in the presence of a nonlocal coupling between the oscillators. This has given rise to a general notion that a {\it weak coupling} approximation (implying phase only oscillators) and {\it nonlocal} coupling are two essential ingredients for the existence of a {\it chimera} state. In a recent work \cite{sethia13} we have demonstrated that the weak coupling approximation is not critical and a more generalized version of the {\it chimera} state that includes amplitude effects can be a collective state of the nonlocal complex Ginzburg-Landau equation (NLCGLE). These amplitude-mediated chimeras (AMCs) can exist as stationary or travelling patterns and show intermittent emergence and decay of amplitude dips in the phase incoherent regions. The next question that naturally arises is whether the nonlocality in the coupling can also be relaxed and whether {\it chimera} states can form through other forms of coupling in a system of oscillators. In this paper we address this issue and show for the first time that the {\it amplitude-mediated chimera} state can emerge even in a globally coupled system of oscillators. We discover these states from a numerical solution of the globally coupled complex Ginzburg-Landau equation (GC-CGLE), a system that has been much studied in the past 
\cite{hakim92,nakagawa93,nakagawa94,hakim94,nakagawa95,chabanol97a,chabanol97c,banaji99,banaji02,
takeuchi09,takeuchi13}  
%\cite{hakim92,nakagawa93,nakagawa94,chate94,hakim94,nakagawa95,chabanol97a,kaneko90,kaneko91,kaneko92,shraiman92,takeuchi13,takeuchi09,takeuchi11,banaji99,banaji02,bragard00,aranson02} 
but none of them reported their existence.  The dynamical nature and origin of these AMCs can be understood from a reduced model description of the GC-CGLE as a forced single oscillator equation with the mean field acting as a periodic driver. Such a description was adopted earlier to explain the dynamics of collective chaos in the GC-CGLE \cite{chabanol97a}. 
It  was shown earlier that the merger of three cluster states (fixed points) in the phase space gives rise to a chaotic state \cite{nakagawa93}. 
From our analysis we find that the AMC state arises from the  coexistence of one fixed point and a limit cycle attractor or a spiral attractor.  A part of the oscillator population that lands up on the fixed point constitutes the coherent portion of the AMC while the rest that migrate to the limit cycle (or spiral attractor) exhibit incoherence under the influence of the mean field fluctuations. The distribution of the oscillators in the complex plane of the amplitude of the oscillators also shows a characteristic signature in the form of a fixed point (for a single coherent cluster) and an extended string like object representing the incoherent portion. 
Our results thus not only expand and extend the concept of {\it chimera} states to globally coupled systems but also provide valuable clues on their dynamical origins that may help us investigate such open questions as the nature of the basins of attraction associated with {\it chimeras} and the stability of these states.  
\begin{figure}
\includegraphics[width = 0.45 \textwidth]{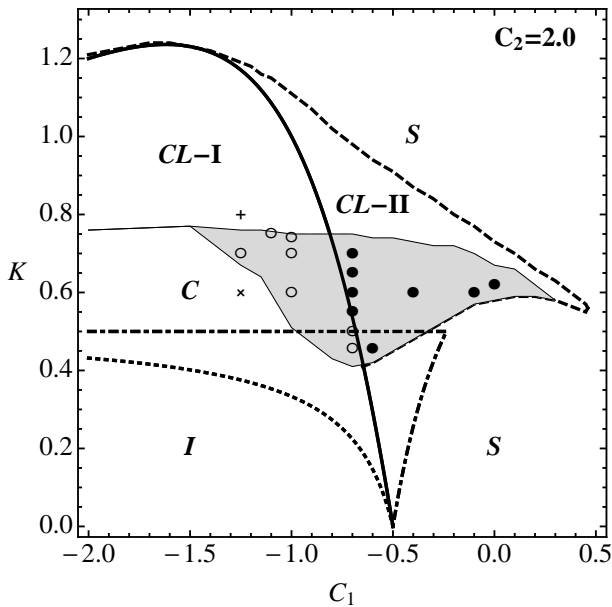} 
\caption{Phase diagram in the $C_1-K$ space for $C_2=2$. The entire region $S$ to the right of the thick solid curve supports stable synchronous (1-cluster) states. Region $I$ below the dotted curve, supports stable splay states while the dot-dashed curve shows the extended limit of the stability of the incoherent states which have nonuniformly distributed phases on the circle. $CL-$I and $CL-$II are regions where multi-cluster states exist with those in $CL-$II co-existing with the synchronous state. Region $C$ has chaotic states while the grey shaded area is  where AMC states exist. Filled circles show typical AMC states coexistent with synchronous states while empty circles are for AMCs in the unstable region of synchronous states. The ($+$) and ($\times$) symbols show positions of a typical 3-cluster state and a chaotic state respectively (see Fig.\ref{fig:fig3}).}
\label{fig:fig1}
\end{figure}

We consider a large population of globally coupled identical oscillators of the complex Ginzburg-Landau type whose dynamics can be represented
by the following set of equations,
\beq 
%\frac{\partial W_{j}}{\partial t} = W_{j}-(1+i C_2)|W_{j}|^{2}W_{j}+ K (1+i C_1) (\overline{W}-W_{j})
\dot{ W_{j}} = W_{j}-(1+i C_2)|W_{j}|^{2}W_{j}+ K (1+i C_1) (\overline{W}-W_{j})
\label{eq:cgle}
\eeq 
where $j=1..N$, with $N$ being the number of oscillators and  $\overline{W}=\frac{1}{N}\sum_{n=1}^N W_n$ is the mean field.
Here $W_{j}(t) $ is a complex field quantity  and $C_1$, $C_2$, $K$ are real constants.
% characterizing the linear dispersion, the non-linear dispersion and the coupling strength respectively.  
Eq.(\ref{eq:cgle}) has a rich dynamics and has been extensively studied in the past \cite{hakim92,nakagawa93,nakagawa94,hakim94,nakagawa95,chabanol97a,chabanol97c,banaji99,
banaji02,takeuchi09,takeuchi13}  
%\cite{hakim92,nakagawa93,nakagawa94,chate94,hakim94,nakagawa95,chabanol97a,kaneko90,kaneko91,kaneko92,shraiman92,takeuchi09,takeuchi11,takeuchi13,banaji99,banaji02,
%bragard00,chabanol97a,aranson02}  
to explore various collective states that occur in the parametric space of $C_1$, $C_2$ and $K$.  A summary of these findings\cite{hakim92,nakagawa93,nakagawa94,chabanol97a}, including ours, is depicted in Fig.\ref{fig:fig1} which is a phase diagram in the $K\;-\;C_1$ space with $C_2=2.0$. The region marked $S$ (comprising of the entire region to the right of the thick solid curve) denotes the existence domain of the fully synchronous state where all oscillators have identical amplitudes and phase and show limit cycle oscillations. The region marked $I$ is the stability domain of the incoherent state where 
the phases of the oscillators are uniformly distributed on a circle (also known as a ``splay'' state \cite{strogatz93}) such that the amplitude of the mean field vanishes. One can also have an incoherent state with a nonuniform distribution of phases such that the mean field still vanishes and the stability region for such states is somewhat larger than $I$ and extends to the boundary marked by a dash-dot curve. $CL-$I and $CL-$II indicate regions where the oscillators break up into two or three coherent clusters where the members within each cluster have identical amplitudes and phases  but these are different for different clusters. Region $C$ marks the domain of the chaotic state
where the dynamical behaviour of each oscillator is quite complex and yet maintains a degree of coherence such that the mean field is still finite. The shaded region in Fig.\ref{fig:fig1} shows the existence domain of the new collective state for this system, namely the {\it amplitude-mediated chimera} state. As seen from the diagram, the existence region of the AMC state includes portions of the phase space that are in the stable region of the synchronous state (implying co-existence) as well as parts of the unstable region to the left of the thick solid curve.

\begin{figure*}
\raisebox{3cm} {(a)}\includegraphics[width = 0.29\textwidth]{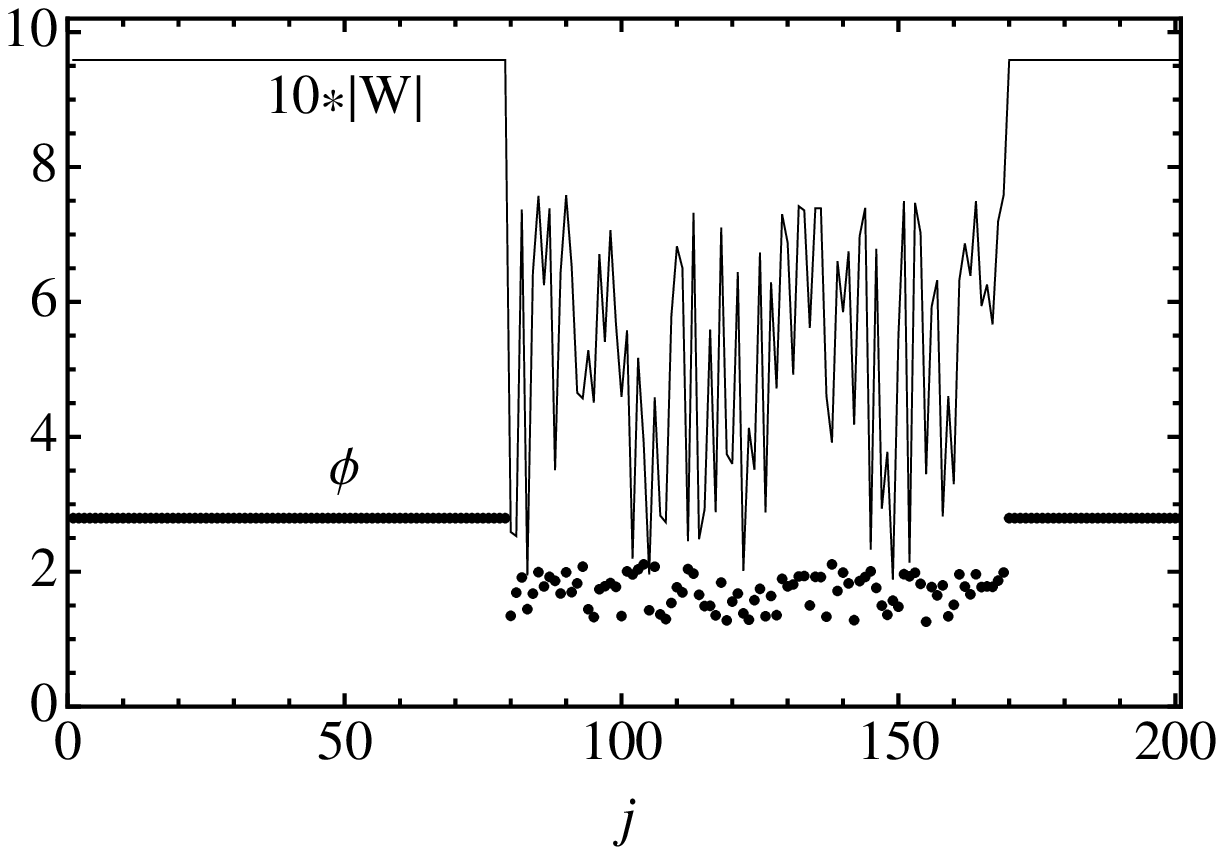} 
\raisebox{3cm} {(b)}\includegraphics[width = 0.29\textwidth]{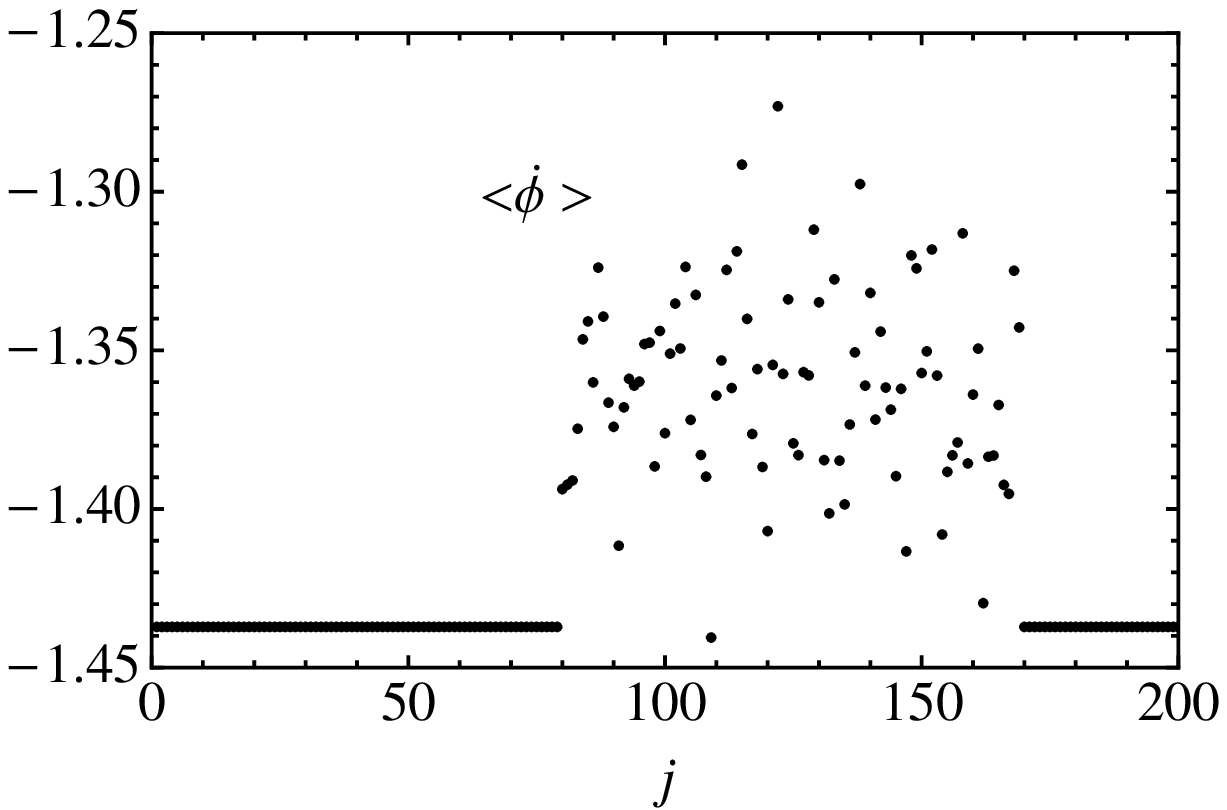} 
\raisebox{3cm} {(c)}\includegraphics[width = 0.29\textwidth]{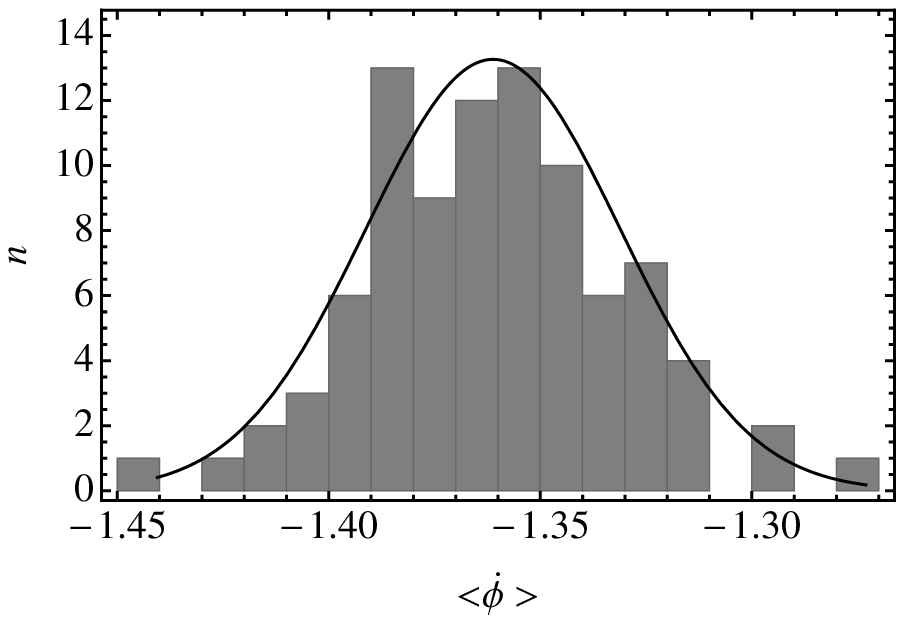} 
\caption{(a) Snapshots of the profiles of the phase ($\phi$) and amplitude $|W|$ (multiplied by 10 ) of an AMC with $K=0.70$, $C_1=-1$ and $C_2=2.0$. The ICs consist  of a splay state. (b) Long-time average of the frequencies ($\dot\phi$) of the oscillators. (c) A histogram of the frequencies ($\langle\dot{\phi}\rangle$) in the incoherent segment with a corresponding Gaussian distribution curve.}
\label{fig:fig2}
\end{figure*}

Our simulations have been carried out with the XPPAUT \cite{xpp} package with a minimum of 201 globally coupled discrete oscillators and the results have been confirmed to remain unchanged with larger number of oscillators. The initial conditions for most of the simulations have been a splay state which is unstable in the parameter space of our simulations. However to check the stability of the AMC state to perturbations in the initial state we have also carried out simulations with initial conditions where the phases have a random distribution. Since under global coupling the index of an oscillator does not mark any spatial location, the final state can always be reconfigured by re-indexing the oscillators. Exploiting this property we have confirmed that the AMC state is indeed stable to perturbations in the initial conditions.  Global coupling also implies no boundaries and hence unlike previous `non-local' {\it chimera} states the AMC states discussed here do not have any dependence on boundary conditions.  A typical portrait of an AMC is shown in Fig.\ref{fig:fig2} where we have displayed the profiles of the amplitude and phases of the oscillators in Fig.\ref{fig:fig2}(a). In Fig.\ref{fig:fig2}(b) we plot the time averaged frequencies of the oscillators which clearly show the breaking up into two sub-groups - those that have a constant frequency (coherent part)  and those that have a distribution of frequencies (incoherent part). In Fig.\ref{fig:fig2}(c) we have plotted a histogram of the frequencies found in the incoherent region that shows a Gaussian profile. 
%Note that the index of an oscillator does not mark any spatial location and the configuration shown in the figure can change as a function of the initial conditions.  
%\begin{figure}
%\raisebox{4.5cm} {(a)}\includegraphics[width = 0.4\textwidth]{Fig3a.eps} 
%\raisebox{4.5cm} {(b)}\includegraphics[width = 0.4 \textwidth]{Fig3b.eps} 
%\caption{(a) The time evolution of the real part of $\overline{W}$ along with the magnitude of $\overline{W}$ are shown. (b) Power spectrum of the real part of $\overline{W}$. The parameter values are as in Fig.\ref{fig:fig2}. }
%\label{fig:fig3}
%\end{figure}

\begin{figure*}
\raisebox{2.8cm} {(a)}\includegraphics[width = 0.3\textwidth]{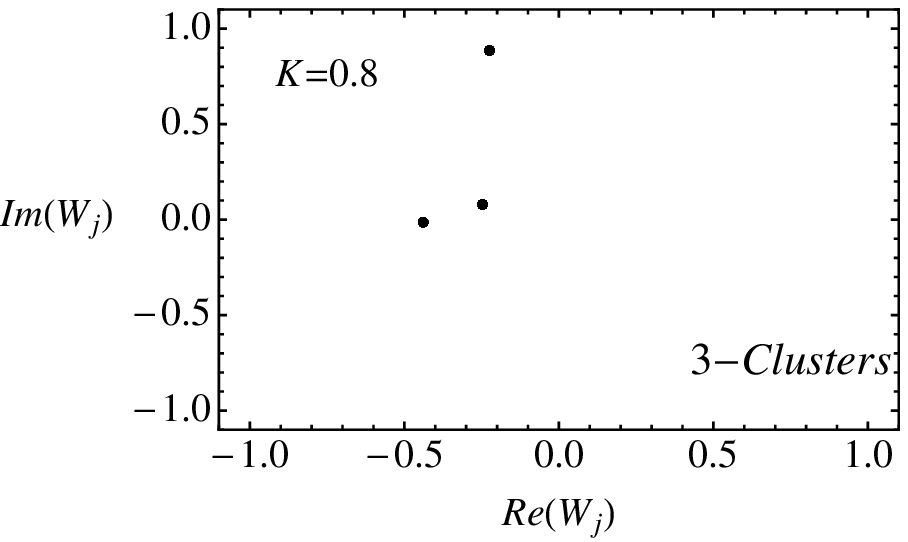} 
\raisebox{2.8cm} {(b)}\includegraphics[width = 0.3\textwidth]{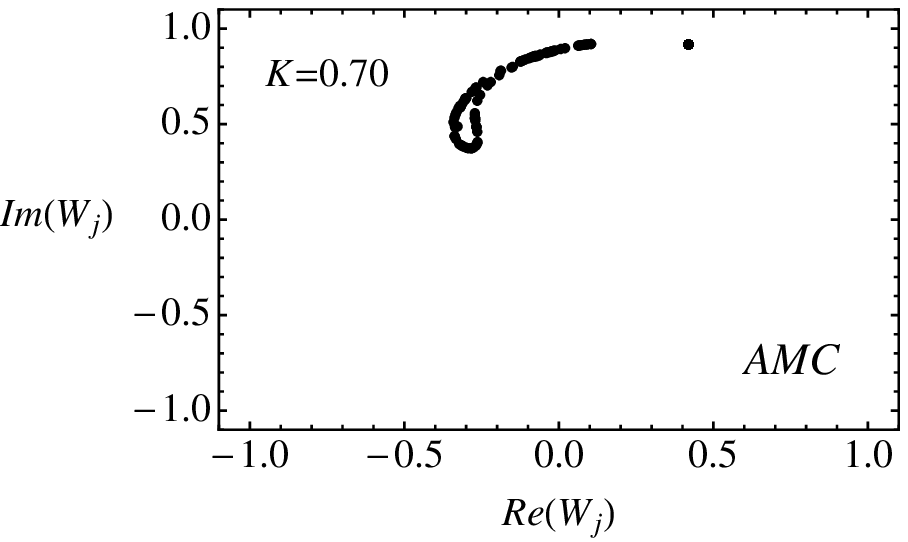} 
\raisebox{2.8cm} {(c)}\includegraphics[width = 0.3\textwidth]{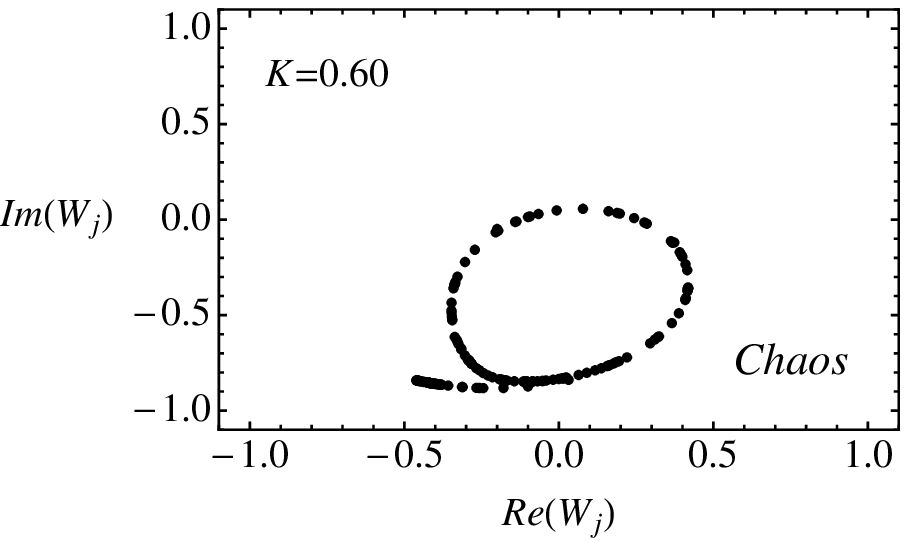} 
\caption{Snapshots of the distribution of 201 oscillators in the complex plane for different values of $K$  with $C_1=-1.25$ and $C_2=2$ corresponding to the leftmost three points in Fig.\ref{fig:fig1}.  As $K$ decreases, two of the three clusters shown in $(a)$ merge to form an AMC state in $(b)$. (c) A further decrease in $K$ leads to merging of all the three clusters leading to chaos. }
\label{fig:fig3}
\end{figure*}

\begin{figure}
\raisebox{2.cm} {(a)}\includegraphics[width = 0.21\textwidth]{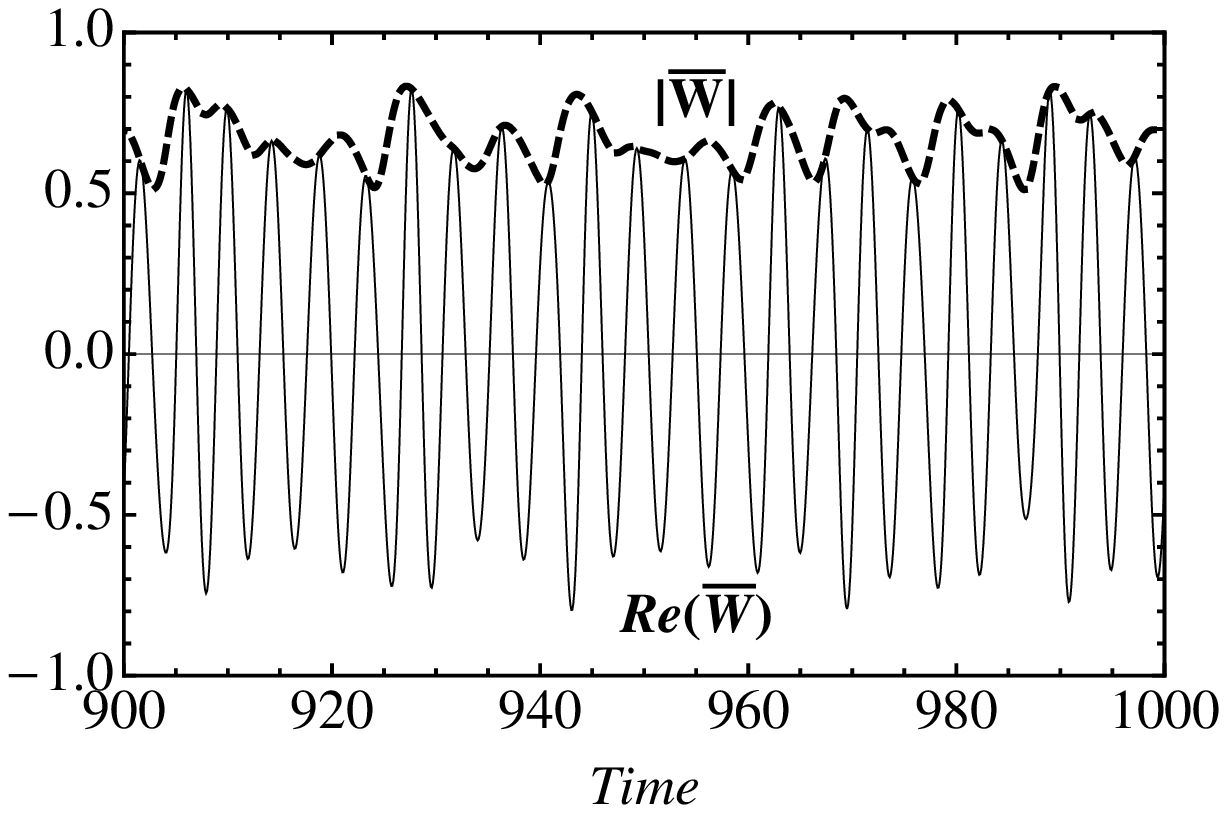} 
\raisebox{2.cm} {(b)}\includegraphics[width = 0.21 \textwidth]{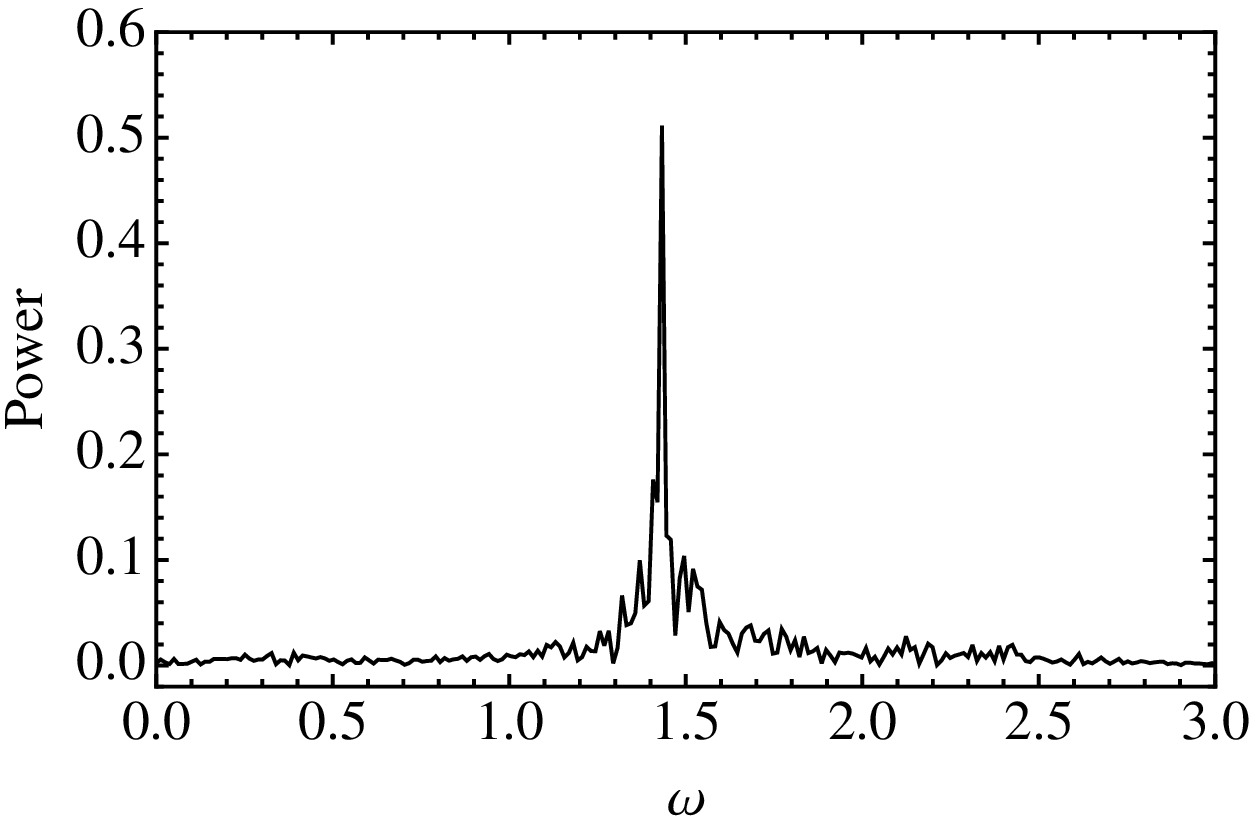} 
\caption{(a) Time evolution of the real part of $\overline{W}$ along with the magnitude of $\overline{W}$. (b) Power spectrum of the real part of $\overline{W}$. The parameter values are as in Fig.\ref{fig:fig2}. }
\label{fig:fig4}
\end{figure}

To gain insight into the dynamical origins of the differences in the behaviour of the different collective states it is instructive to plot the position of each oscillator in the complex plane of $Im (W_j)$ versus $Re( W_j)$. In Fig.\ref{fig:fig3} we compare the typical temporal snapshots of the cluster state, the AMC and the chaotic state. Fig. \ref{fig:fig3}(a) shows a three cluster state, where the three clusters rotate as a function of time maintaining their individual sizes. The chaotic state shown in Fig.\ref{fig:fig3}(c) happens when all three clusters merge to give rise to a $\rho$ shaped distribution that rotates as well as distorts as a function of time. The AMC emerges from an intermediate configuration where two of the clusters merge to produce a string like object and one cluster retains its separate identity as shown in Fig.\ref{fig:fig3}(b). As a point of historical interest, we would like to mention that the configuration of Fig.\ref{fig:fig3}(b) was noticed by Nakagawa and Kuramoto \cite{nakagawa93} while exploring the chaotic state of Eq.(\ref{eq:cgle}) but set aside as something where the ``nature of chaos is not clear yet''.  This was nearly a decade before the discovery of the {\it chimera} state by Battogtokh and Kuramoto \cite{kuramoto02}. The oscillators residing in the cluster constitute the coherent portion whereas those on the string are the incoherent portion of the AMC. This transition between the different states is governed by the values of the parameter $K$,$C_1$ and $C_2$. As an illustration we have marked the positions of the three states whose snapshots are shown in Fig.\ref{fig:fig3} with different symbols in Fig.\ref{fig:fig1}. The plus symbol marks the position of the three cluster state. As we then descend to lower values of $K$ for a fixed value of $C_1=-1.25$ we enter the regime of AMCs and the position of one of them is shown by the symbol of an open circle. Decreasing $K$ further brings us to the chaotic regime with the cross symbol marking the position of one such representative state. A similar scenario holds for the emergence of the AMC in the stable region of the synchronous state except that with the decrease of $K$, the AMC does not go to a chaotic state but reverts to a synchronous state. 

To gain further understanding of the dynamics of the AMCs,  we next analyze a reduced model of the GC-CGLE that has been previously employed \cite{chabanol97a} to study the dynamics of the  chaotic state. We define the mean field as  
$\overline{W}(t)=Re^{i\omega t}$, where $R$ (assumed constant) is the amplitude of the mean field and $\omega$ is a frequency around which the real part of the mean field is peaked. Eq.(\ref{eq:cgle}) can then be reduced to:
\beq 
\frac{\partial B}{\partial \tau} = (1+i\Omega)B-(1+i C_2)|B|^2B+F
\label{eq:rcgle}
\eeq 
where $B(t)$ is the transformed function in a rotating frame to remove explicit time dependence and is given as 
%\beq
$B(t)=(1-K)^{-1/2}W(t)e^{-i(\omega t+\phi)}$
%\eeq
with
\bea
\tau=(1-K)t\nonumber\\
\Omega=-\dfrac{\omega+K C_1}{1-K}\nonumber\\
F=\dfrac{K\sqrt{1+C_1^2}}{(1-K)^{3/2}}R
\label{eq:of2}
\eea
where $\tau$ is a rescaled time, $C_1=\tan(\phi)$ and $-\pi/2<\phi<\pi/2$.
\begin{figure}
\includegraphics[width = 0.45 \textwidth]{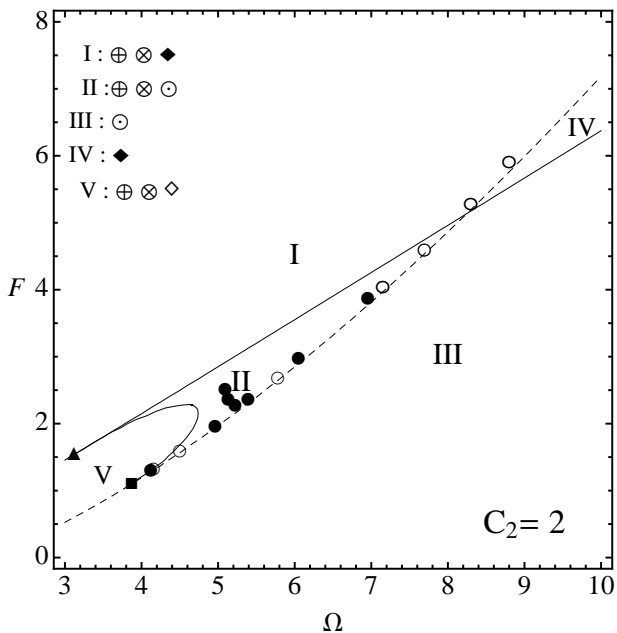} 
\caption{ Bifurcation diagram (adapted from \cite{chabanol97a}) of Eq.(\ref{eq:rcgle}) in $F- \Omega$ space, with the Hopf bifurcation line indicated by a solid curve and a dashed curve showing the saddle-node bifurcation line. The thin line between the regions $II$ and $V$ is the homoclinic bifurcation line. The filled triangle and the square points are the Takens-Bogdanov bifurcation point and the co-dimension-two points respectively. The filled and the empty circle markers correspond to the AMC points shown in Fig.\ref{fig:fig1}.  In the legend at the top-left corner of the figure, the symbols circle -plus, -times and -dot denote a node, a saddle and an attractive limit cycle respectively. The filled- and the empty- diamonds denote stable and unstable spirals respectively.}
\label{fig:fig5}
\end{figure}
The assumptions underlying the simplification adopted above can be justified by looking at the actual forms of $|\overline{W}|, Re(\overline{W})$ and its power spectrum as obtained from the numerical solution of Eq.(\ref{eq:cgle}) and shown in Fig.\ref{fig:fig4}(a,b). It is seen that the time evolution of the real part of $\overline{W}$ is nearly periodic around a single frequency denoted by the peak of the power spectrum. The mean amplitude of $\overline{W}$ is also seen to vary on a slower time scale than $\omega$ and hence is taken to be nearly constant.  Thus the entity $F$ serves as a periodic driver for each oscillator and the dynamics of the system can be understood in a simple manner from the evolution of this single driven oscillator equation. We make use of the detailed bifurcation diagram of Eq.(\ref{eq:rcgle}) that has been obtained in \cite{chabanol97a} and that is shown in Fig.\ref{fig:fig5} to understand the dynamics of the AMC state.  

In the centre region $II$ of the  bifurcation diagram, the phase portrait consists of an attractive limit cycle  with a unstable spiral point within it, and a pair of two other fixed points: one of which is a node and the other one is a saddle point. As any point within this region gets closer to the Hopf bifurcation line, the limit cycle shrinks around its inside unstable spiral point and above the Hopf bifurcation line (region $I$), it is replaced by a stable spiral point.  The saddle point and the node  in region $II$ collide and disappear on the saddle-node bifurcation line leaving behind an attractive limit cycle in region $III$ and a stable spiral point in region $IV$. As one moves from region $II$  to the centre-left region $V$, the saddle point outside the limit cycle moves towards the limit cycle and transforms it into a homo-clinic orbit on the thin bifurcation line shown between these two regions. The three fixed points (a node, an unstable saddle and an unstable spiral) remain in region $V$. As one crosses the Hopf bifurcation line from the region $III$ to the region $IV$, the unstable spiral point within the limit cycle gets replaced by a stable spiral point. We have marked the parameter values of the AMC states shown in Fig. \ref{fig:fig1}  by filled and empty circles on the bifurcation diagram. As can be seen they closely
hug the saddle node (SN) bifurcation curve indicated by the dashed line. The AMC state is governed by the topology of the regions I and II close to the SN curve,  where there is
coexistence of a stable node and a  limit cycle or a spiral attractor.  In the full system the fluctuations in the amplitude of the mean field which act as perturbations to the periodic driver of the reduced system drive oscillators towards these equilibrium points. The oscillators that go to the node remain there and constitute the coherent part of the AMC while those that populate the limit cycle or the stable spiral make up the incoherent part of the AMC. 
The distribution of the oscillators among these two sub-populations depends on the initial conditions and the kicks in phase space that they receive from the amplitude fluctuations. Thus finite amplitude effects (a consequence of the strong coupling limit) play an important role in the formation of the AMCs and the additional degree
of freedom that they introduce in the system appears to allow one to do away with the nonlocal coupling. Our present model is in some sense a generalization of the minimal model consisting of two interacting populations of phase oscillators that was proposed and investigated in \cite{abrams08} . In that model each phase oscillator was coupled equally to all other oscillators within its own group and less strongly coupled to all oscillators in the other group and the difference in the strength of the two interactions was necessary to simulate the effect of a nonlocal interaction in order to obtain a {\it chimera} state. In our model the coupling is uniform for all oscillators and the amplitude effects contribute towards the formation of the collective AMC state.

Our results therefore open up a much broader framework for the emergence of {\it chimera} states by freeing it from the constraints of the weak coupling approximation and the necessity of a nonlocal coupling among oscillators.  In particular, the existence of an AMC state in a globally coupled system, one of the most widely used dynamical systems model, can have significant implications for practical applications in a variety of physical, chemical and biological phenomena. As an example, the global model with a source term is often used in deep brain stimulation studies \cite{popovych05} to determine effective strategies for inducing de-coherence in the region affected by Parkinson's disease.   Our findings about the dynamical origins of the AMC and their signatures in phase space could be valuable in further refinement of such techniques as well as stimulate basic research on such  open problems as the stability of these states and determination of their basins of attraction in parameter space.

Acknowledgement: G.C.S. acknowledges the support of MPI-PKS, Germany, where the present work got initiated. He also acknowledges his valuable discussions with Prof.~Hugues Chat$\acute{e}$ while at MPI-PKS and with Prof.~Vincent Hakim through email exchanges.
\bibliographystyle{apsrev}
%\bibliographystyle{unsrt}
%\bibliography{refs_Feb_2014}
%%%%

\end{document}